\documentclass[twocolumn]{revtex4-1}
\usepackage[applemac]{inputenc}
\usepackage{graphicx}
\usepackage{color}
\usepackage{amsmath}
\usepackage{lineno}
 \usepackage{relsize}
 \DeclareMathSizes{17}{10}{5}{6}
\newcommand{\bea}{\begin{eqnarray}}
\newcommand{\eea}{\end{eqnarray}}
\newcommand{\beq}{\begin{equation}}
\newcommand{\eeq}{\end{equation}}
\begin{document}
\title{Sensitivity of multiplicity fluctuations to freeze-out conditions in heavy~ion collisions}
\author{Paolo Alba$^a$, Rene Bellwied$^b$, Marcus Bluhm$^c$}
\author{Valentina Mantovani Sarti$^{a}$, Marlene Nahrgang$^d$, Claudia Ratti$^{b}$}
\address{$^a$ Department of Physics, Torino University and INFN, Sezione di Torino, via P. Giuria 1, 10125 Torino, Italy\\
$^b$ Department of Physics, University of Houston, Houston, TX 77204, USA\\
$^c$ Department of Physics, North Carolina State University, Raleigh, NC 27695, USA\\
$^d$ Department of Physics, Duke University, Durham, NC 27708-0305, USA}
\begin{abstract}
We study the sensitivity of the higher-order moments of produced particle multiplicity distributions to the chemical freeze-out parameters in relativistic heavy ion collisions using the Hadron Resonance Gas (HRG) model. We compare the obtained sensitivity level to the one extracted from the ratios of particle yields. We find that, for certain final state hadrons, the fluctuation measurements add significant information to the determination of the hadro-chemical freeze-out properties of the deconfined phase of matter obtained at RHIC and the LHC.
\end{abstract}

\maketitle

\section*{Introduction}

Over the past few decades the particle production in relativistic heavy-ion collisions has been successfully described by
statistical hadronization models which postulate a chemical freeze-out surface common to all hadronic species. This means the yield of particles emitted from the interaction region can be reproduced by assuming an equilibrated system with fixed thermodynamic parameters. At this time in the dynamical evolution of the expanding fireball, the production of particles through inelastic collisions ceases and freeze-out of the hadro-chemistry is achieved. The resulting phase space surface is generally characterized (at sufficiently high collision energies) by only two free parameters, namely the chemical freeze-out temperature and the associated baryo-chemical potential ($T_{ch}$ and $\mu_{B,ch}$). These types of static models have proven to be extremely successful over a wide range of collision energies, which led to the definition of a generally accepted chemical freeze-out curve in the $T-\mu_{B}$ plane~\cite{Cleymans:2005xv}. It was also shown that this curve is close to the QCD transition curve as defined by lattice QCD~\cite{Tctrilogy}. Since the QCD transition is an analytic crossover 
for small $\mu_{B}$~\cite{Aoki:2006we}, the lattice curve corresponds to the steepest gradient in the temperature dependence of some characteristic observables which identify the deconfinement and chiral restoration transition~\cite{Tctrilogy,Bazavov:2011nk}.

In the past, statistical model calculations were used only to describe the multiplicity of particles, either by calculating the expected yields (which adds the interaction volume as an additional free parameter to the calculation) or by calculating particle ratios (in which case the volume cancels out, assuming a common, flavor independent freeze-out volume)~\cite{Cleymans:2005xv,BraunMunzinger:2003zd,Becattini:2003wp,Becattini:2005xt,Andronic:2011yq,Sagun:2014sya}.

Recently, it has been shown that the moments of net-particle multiplicity distributions from the experiment can be related to susceptibilities of conserved charges calculated on the lattice~\cite{Karsch:2012wm,Bazavov:2012vg,Borsanyi:2013hza,Borsanyi:2014ewa}. This allows the direct determination of chemical freeze-out parameters in the thermally equilibrated grand-canonical ensemble approach on the lattice without having to rely on statistical models.

The continued application of statistical models in the context of these lattice QCD-to-experiment comparisons, is based on the fact that conserved charges on the lattice can only be directly related to particle distribution moment measurements by studying and controlling the limitations of the experiment to measure conserved charges. Thus, by modeling detector effects, such as acceptance and reconstruction efficiency, in a statistical model calculation, the latter de-facto facilitates the necessary link between experiment and lattice QCD. 
It was shown that statistical Hadron Resonance Gas (HRG) models reproduce the equilibrium lattice QCD results for the lowest order susceptibilities and their ratios in the hadronic phase reasonably well~\cite{Borsanyi:2013hza,Borsanyi:2014ewa,Mukherjee:2013lsa}. An added advantage of the HRG model calculations is the fact that a conserved charge, such as strangeness, can be broken down into its contributions from each identified particle species, which is not achievable on the lattice.

We have shown previously that the calculations of the lowest moments of particle multiplicity distributions in a statistical model are in agreement with recent STAR data~\cite{Alba:2014eba}. One can therefore ask whether the chemical freeze-out parameters obtained from our analysis are comparable to parameters from an analysis of the particle multiplicities. For this purpose, we study here the sensitivity of higher-order moment ratios on the freeze-out parameters in comparison to that of multiplicity ratios. Depending on the observable and the experimentally achievable accuracy in the corresponding measurement, it might be advantageous to consider fluctuations rather than yields for pinning down the chemical freeze-out conditions. This requires that the measured fluctuations are those of an equilibrated hadronic medium~\cite{Kitazawa:2013bta,Sakaida:2014pya}.

From the experimental data the central moments of the (net-)particle distributions can be constructed. 
The corresponding cumulants are given by:

{\footnotesize
\bea
c_1&=&\langle
(N)\rangle~~~~~~~~~~~~~~~~~~c_2=\langle\left(\delta
N\right)^2\rangle
\nonumber
\\
&&
\nonumber
\\
c_3&=&\langle\left(\delta
N\right)^3\rangle~~~~~~~~~~~~~~~~c_4=\langle\left(\delta
N\right)^4\rangle-
3\langle\left(\delta N\right)^2\rangle^2
\nonumber
\eea}
where $\delta N=N-\langle N\rangle$ is the fluctuation of the (net-)particle multiplicity around its 
mean. The cumulants $c_i$ relate directly to the susceptibilities $\chi_i$, which are quantities that 
can be calculated for thermodynamic systems, e.g.~in lattice QCD. Susceptibilities are defined as derivatives of the pressure with respect to the chemical potential. In order to quantify certain features of distributions beyond the mean ($M=\chi_1$) and the variance ($\sigma^2=\chi_2$) one often looks at the skewness $S$ and the kurtosis $\kappa$ defined as:

{\footnotesize
\bea
\mathrm{ mean:}~M=\chi_1~&&~\mathrm{
variance:}~\sigma^2=\chi_2
\nonumber
\\
&&
\nonumber
\\
\mathrm{
skewness:}~S=\chi_3/\chi_{2}^{3/2}~&&~\mathrm{kurtosis:}~\kappa=\chi_4/\chi_{2}^{2}
\nonumber
\eea}

Then, one can relate the thermodynamic susceptibilities calculated on the lattice or in an HRG model 
to quantities obtained experimentally through the following volume-independent ratios:

{\footnotesize
\bea
S\sigma=\chi_3/\chi_{2}~~~~~~&&~~~~~~\kappa\sigma^2=\chi_4/\chi_{2}
\nonumber\\
&&
\nonumber\\
M/\sigma^2=\chi_1/\chi_{2}~~~~~~&&~~~~~~S\sigma^3/M=\chi_3/\chi_{1}.
\nonumber
\eea}

\section*{Details of the Hadron Resonance Gas model}

Detailed measurements of conserved charge fluctuations have been conducted at several collision energies at RHIC in order to search for non-statistical fluctuations that could signal the existence of a critical point in the QCD phase diagram~\cite{Adamczyk:2013dal,charge,McDonald:2012ts}. A series of studies using variations of the standard statistical hadronization models to determine the baseline for these critical point searches can be found in the literature ~\cite{Begun:2006jf,Karsch:2010ck,Fu,Garg:2013ata,Nahrgang:2014fza}.
The results presented in this paper are obtained using an HRG model in partial chemical equilibrium, which means that all contributions from the strong decay of hadronic resonances are taken into account. The list of all included resonant states is based on the PDG tables \cite{PDG12}, up to a mass of 2 GeV/c$^{2}$. 

Further details of this HRG model can be found in~\cite{Alba:2014eba}, where our group compared the net-charge and net-proton distribution measurements from the STAR collaboration to the HRG model results.
We evaluated ratios of the lower moments of these multiplicity distributions as functions of the temperature $T$, baryo-chemical potential
$\mu_B$ and the chemical potentials $\mu_Q$ and $\mu_S$. The relation among these thermodynamic variables is obtained by imposing certain initial conditions occurring at the collision, namely 
the conservation of the total net-strangeness $n_S=0$ and the proper ratio of protons to baryons in the colliding nuclei $n_Q=\tfrac{Z}{A}n_B$.

 We obtained lower freeze-out temperatures ($\approx$146 MeV) than the statistical hadronization model fits to all available particle multiplicities ($\approx$166 MeV). We attributed this discrepancy in part to the fact that net-proton and net-charge are mainly sensitive to the light quark susceptibilities, whereas a full fit to all particle multiplicities contains a significant contribution from (multi-)strange particles. A possible separation of light and strange
quark transitions was suggested by high precision lattice
QCD simulations of the kurtosis ~\cite{Bellwied:2013cta}. If true, then strange particle multiplicities and moments might yield a higher freeze-out temperature, as already pointed out in~\cite{Alba:2013haa,Bluhm:2014lva}.

 A comparison of our freeze-out temperature obtained from the net-proton and net-charge distribution to the one from measured particle yields (see Ref. \cite{Alba:2014eba}), indeed indicates a discrepancy between the temperature necessary to describe the strange and multi-strange baryon yields compared to the lower mass and mostly light quark states ($\pi,\,K,\,p$). In this fit to the particle multiplicities pions and kaons, though, show very little sensitivity to the chosen freeze-out temperature at fixed chemical potential. Since pions, kaons and protons are presently the only particles for which the experiments can determine higher order fluctuations, it is interesting to ask whether, for these hadronic final states, the moments of the net distributions are more sensitive to the chemical freeze-out conditions than the basic particle ratios.

Our initial study has also shown that the highest cumulants might be prone to non-statistical effects such as volume fluctuations or chiral criticality, even at collision energies in the crossover region~\cite{Friman:2011pf}. Critical behavior might be captured in lattice QCD, but statistical models are by definition insensitive to these dynamical fluctuations. Therefore we initially focus our study on the lowest moments necessary to determine freeze-out parameters, namely $\chi_2$ and $\chi_1$. Whether $\chi_3$ already contains a small contribution from critical fluctuations is an open issue, which is presently debated in the literature. For very high collision energies, well separated from a potential critical point, these effects should be small. Thus, for the highest RHIC energies and all LHC energies it should be safe to also use $\chi_3$ or even $\chi_4$ for a potential determination of the chemical freeze-out parameters, and we show results also for moment ratios that include the higher moments.

\section*{Results}

Based on the above mentioned tension in the extracted chemical freeze-out temperatures using the STAR net-proton and net-charge fluctuation results compared to the particle ratios, in particular at small baryo-chemical potentials~\cite{Alba:2014eba}, we perform the comparative study presented here only for the highest RHIC energy. Since there is good agreement between the baryo-chemical potentials deduced from the particle ratios and the fluctuations, we fixed $\mu_B$ at 24.3 MeV and determined the sensitivity of different observables to the freeze-out temperature. Historically, ratios of mesons and baryons to pions have proven to be a good thermometer, since their dependence on the temperature $T_{ch}$ is mainly driven by the particle mass difference. Particle yields might have a higher sensitivity to the temperature but a fit to ratios is preferred, since the volume cancels out and the ratios are less prone to biases \cite{Andronic:2005yp}. Our results for the sensitivity of particle ratios to the temperature are in good agreement with previous statistical model calculations \cite{Magestro:2001jz,Andronic:2005yp}. Here we add the sensitivity of the moments to this study, and we show that for certain particles it might be beneficial to use the moment analysis, if efficiency corrected data are available.

\begin{figure}[h!]
\centering
\includegraphics[width=0.45\textwidth]{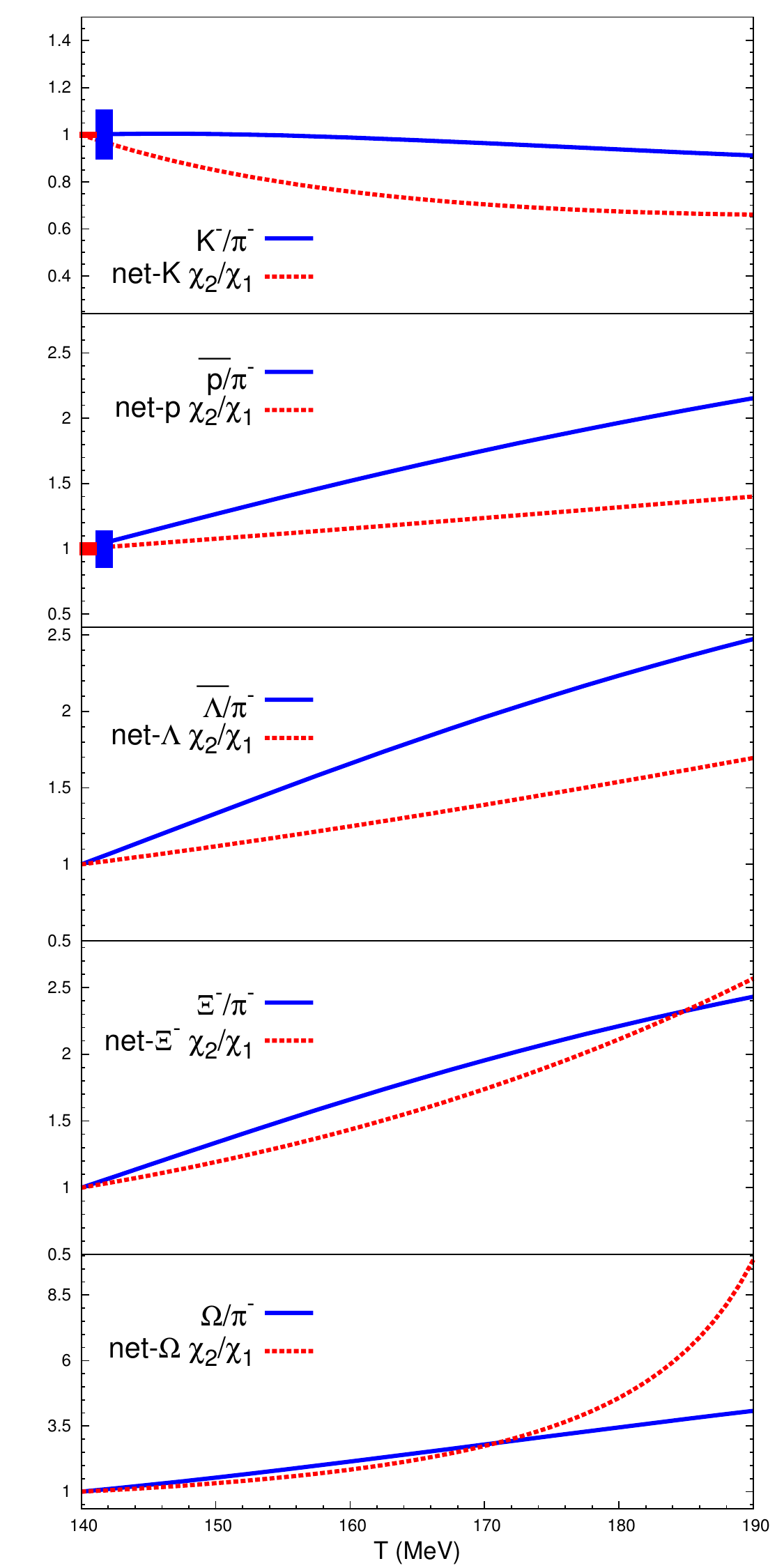}
\caption[]{\label{fig2}(Color online) Comparison of the sensitivity of $\chi_2$/$\chi_1$ ratios and particle yield ratios to the chemical freeze-out temperature. All values are normalized to the value at 140 MeV. The baryo-chemical potential has been fixed to 24.3 MeV. The vertical bars in the plots show the presently achievable experimental uncertainties for particle species for which both particle yield and moment ratios are available (based on STAR measurements~\cite{Adamczyk:2013dal,McDonald:2012ts,Adams:2006ke,Abelev:2008ab}).}
\end{figure}

A comparison of the freeze-out temperature sensitivity of the particle ratio (with respect to the pion yield) and the lowest moment ratio ($\chi_2$/$\chi_1$) for all main hadron species ($K, p, \Lambda$, $\Xi$, $\Omega$) is shown in Fig.~\ref{fig2}. In particular for kaons the difference is striking: the $K/\pi$ ratio is almost completely insensitive to the temperature, whereas the net-kaon moment ratio shows a strong sensitivity well outside the achievable experimental error bars.  

The
experimental error bars, shown as vertical thick lines in Fig.~$1$, are from~\cite{Abelev:2008ab} for the particle ratios, the net-proton
$\chi_2/\chi_1$ uncertainty is from the recent STAR publication~\cite{Adamczyk:2013dal}. For the net-Kaon
$\chi_2/\chi_1$ error we have used a preliminary study by STAR~\cite{McDonald:2012ts} based on moments
that were not efficiency corrected. The effect of the reconstruction efficiency
correction was estimated to roughly double the uncertainty~\cite{McDonald:private} and the error bar shown in Fig.~$1$ reflects the anticipated
error for an efficiency corrected measurement. In general, we show experimental uncertainties only for particle species for which both measurements (the yield ratios and the net-moment ratios) have been published.

It is interesting to note that for protons, which require a lower temperature for the yields and moments at the LHC and the higher RHIC energies, both measurements show the necessary sensitivity based on the experimentally achievable error bars. The lower temperature deduced for protons from lattice QCD and HRG model moment analyses~\cite{Borsanyi:2014ewa,Alba:2014eba}, might thus explain the 'proton anomaly' in the yield. Generally, the baryon yield ratios show a stronger sensitivity to the temperature than the corresponding lower moment ratios; nevertheless, for multi-strange hyperons, both particle yield and moment ratio exhibit a comparable steepness in the relevant temperature range.

\begin{figure}[h!]
\centering
\includegraphics[width=0.45\textwidth]{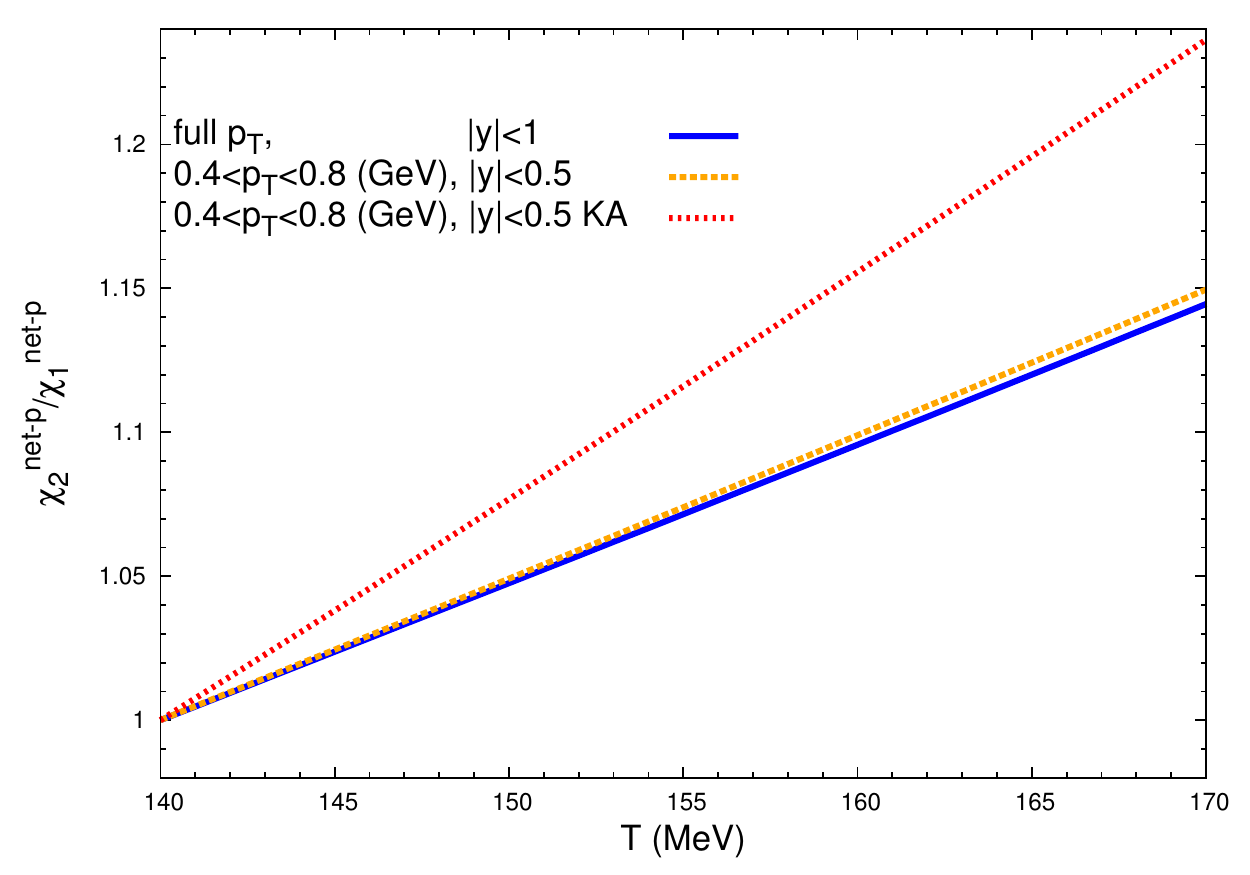}
\caption[]{\label{fig1} (Color online) Net-proton $\chi_2/\chi_1$ as a function of the temperature: comparison between the curves obtained with different kinematic cuts and with/without isospin randomization. All values are normalized to the proper value at 140 MeV.}
\end{figure}

The moment ratios which we show in Fig.~1 include kinematic cuts in rapidity and transverse momentum according to the experimental restrictions as well as contributions from resonance decays. For the latter, two effects have to be taken into account. First, a resonance can decay and be regenerated, which leads, for sufficiently short-lived resonances, to an isospin randomization for the daughter hadrons. This effect is for example relevant for the net-protons since the $\Delta$-resonances have a very short lifetime. Kitazawa and Asakawa ~\cite{Kitazawa:2011wh,Kitazawa:2012at} have proposed corrections, which are applied to the net-proton moment ratio in Fig.~1 (see also~\cite{Nahrgang:2014fza}). The magnitude of the kinematic cuts and the isospin randomization effect  on the slope of $\chi_2/\chi_1$ is shown in Fig.~2 for net-protons: while different cuts on momentum and rapidity have a very small effect~\cite{Garg:2013ata}, the corrections due to isospin randomization (KA corrections) visibly increase the sensitivity of this curve to the temperature. The net-proton $\chi_2/\chi_1$ shown in Fig.~1 corresponds to the red, dotted line in Fig.~2, for which all possible effects are taken into account.

\begin{figure}[h!]
\centering
\includegraphics[width=0.45\textwidth]{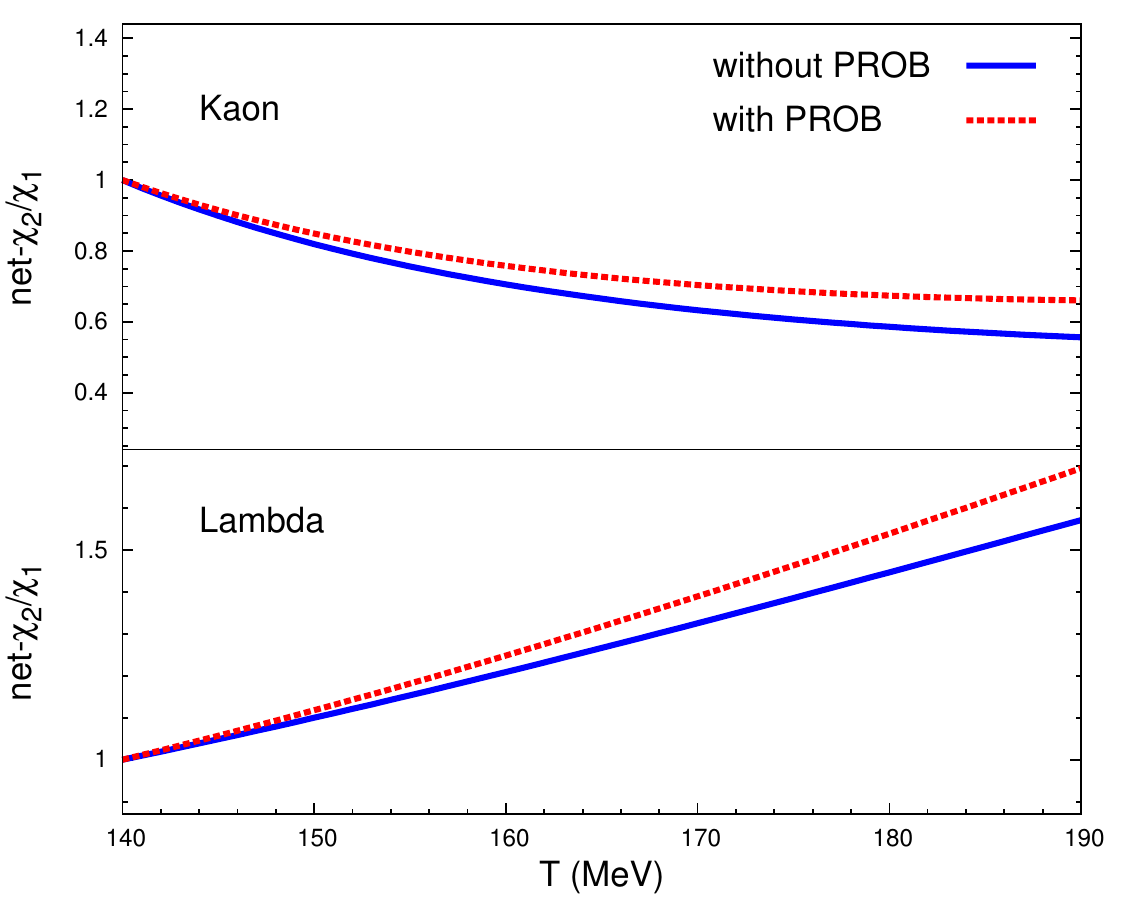}
\caption[]{\label{fig1bis} (Color online) Net-kaon and net-$\Lambda$ $\chi_2/\chi_1$ as a function of the temperature: comparison between the curves with/without the probabilistic decay contribution (PROB). All values are normalized to the proper value at 140 MeV.}
\end{figure}

The second relevant effect is due to the probabilistic nature of the resonance decay, because the number of ground state hadrons follows a probability distribution due to the different branching options. This was first pointed out 
in~\cite{Begun:2006jf} and can be applied to the susceptibilities $\chi_i$ with $i>1$ by taking into account the actual decay contributions rather than an averaged branching pattern~\cite{Fu}. We apply this probabilistic effect to all particles investigated in Fig.~1 except the (anti-)protons as here isospin randomization dominates the probabilistic resonance decay contributions. Fig.~\ref{fig1bis} shows the net-kaon and net-$\Lambda$ $\chi_2/\chi_1$ with and without the probabilistic decay contribution (PROB). As in the case of the isospin corrections, the temperature sensitivity changes when the effect is taken into account. It is interesting to note, though, that the relative temperature sensitivity for net-kaon fluctuations reduces, whereas for net-$\Lambda$s the sensitivity is enhanced.

\begin{figure}[h!]
\centering
\includegraphics[width=0.45\textwidth]{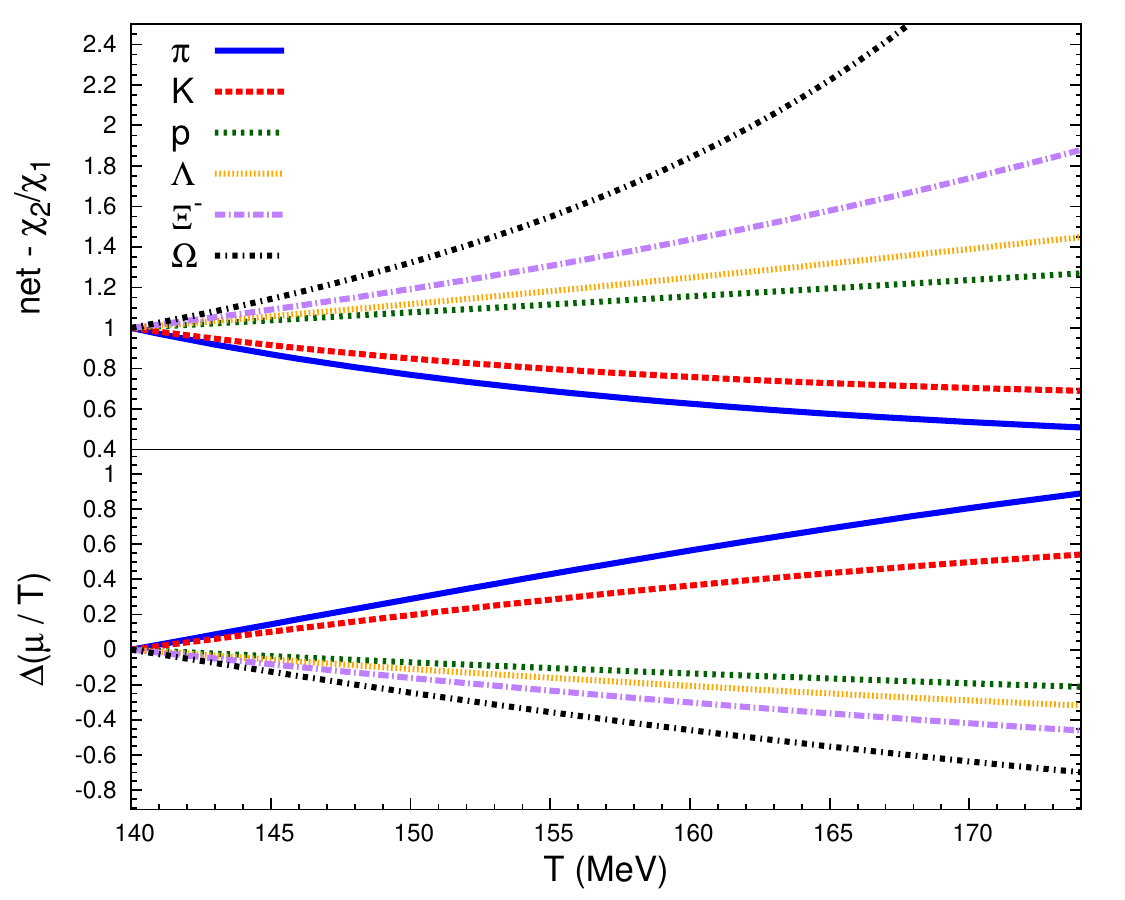}
\caption[]{\label{fig3}(Color online) Upper panel: comparison of the sensitivity of $\tfrac{\chi_2}{\chi_1}$ ratios to the chemical freeze-out temperature for different hadron species.
 Lower panel: $\Delta (\tfrac{\mu_i(T)}{T})$ for particle $i$ as a function of the temperature for mesons and
baryons. All values are normalized to the proper value at 140 MeV and the baryo-chemical potential has been fixed to 24.3 MeV.}
\end{figure}

The upper panel of Fig.\ref{fig3} shows a comparison of the net-$\chi_2$/$\chi_1$ temperature sensitivity for all relevant stable hadrons that are measurable by the experiments. 
A noticeable baryon-to-meson difference can be established. This difference can be understood qualitatively already for primordial hadrons (without resonance decays) in the Boltzmann approximation, where the net-cumulant ratio $\chi_2/\chi_1$ is given by $\coth(\mu_i/T)$, with $\mu_i$ being the chemical potential of particle $i$~\cite{Karsch:2010ck}.
The ratio $\tfrac{\mu_i}{T}$ has a different trend for mesons and baryons, as can be seen in the lower panel of Fig.~\ref{fig3} (note that the normalized difference from the value at $T=140$~MeV is shown), because the individual components that contribute to $\mu_i$, i.e. $\mu_B,\mu_Q,\mu_S$, sum up to an opposite temperature dependence for baryons and mesons.

Let us point out that, due to resonances which can decay into a meson and/or its anti-meson, the susceptibilities for net-mesons cannot simply be obtained via independent production. For the second susceptibility this is seen from:

\begin{align}\label{eq1}
\mathlarger \chi_{2,net-i} = &  \mathlarger\chi_{2} ^i+\mathlarger\chi_{2} ^{\bar{i}}+ \nonumber \\
& \sum  _R (\langle \mathlarger n_i\rangle _R ^2+\langle \mathlarger n_{\bar{i}}\rangle _R ^2
 -2\langle \mathlarger n_i\rangle _R\langle \mathlarger n_{\bar{i}}\rangle _R)\mathlarger\chi_2 ^R .
\end{align}

The first two terms, $\chi_2 ^{i,\bar{i}}$, represent the contributions from the primordial particles, while the sum in Eq.~(\ref{eq1}) accounts for the average resonance decay contributions. The last term in Eq. (\ref{eq1}) is non-zero whenever a resonance decays into a stable particle, its antiparticle or the particle-antiparticle pair. Since there are no known hadronic resonances that decay into a baryon and/or its anti-baryon, Eq.~(\ref{eq1}) reduces to independent production in the case of baryons, while correlations arise for mesons.

It has been established, on the basis of simple particle yields, that in a statistical model the strange baryons are more sensitive to the temperature since the strange quark rest mass is close to the temperature of the equilibrated system.
As shown in Fig. \ref{fig3}, the steepness of the $\chi_2/\chi_1$ curves, and thus the sensitivity to the temperature, 
increases with the strange quark content, which means that cumulant ratios in the multi-strange baryon sector rise 
significantly faster than for the $\Lambda$. This effect is mainly caused by imposing vanishing net-strangeness, 
which means that even at constant $\mu_B$, the strange chemical potential $\mu_S$ has to 
increase as a function of the temperature (in our simulation from 3 to 8 MeV between 140 and 180 MeV). The impact can 
already be seen in the temperature dependence of the $\Lambda$ $\chi_2$/$\chi_1$. For the multi-strange baryons this rise
grows according to the strange quark content, each quark leading to an additive strangeness 
chemical potential contribution. In other words, for the net-$\Omega$ at high $T$ the strangeness chemical potential ($\mu_S$ = 3$\times$8 MeV = 24 MeV), offsets 
the baryo-chemical potential ($\mu_B$ = 24.3 MeV) since the two chemical potentials always carry opposite sign. 
At this point the net-multiplicity grows much slower than the variance, which leads to a strongly
increasing $\chi_2$/$\chi_1$ ratio. This effect is not unique to the specific value of the baryo-chemical potential, in other words at lower
and higher collision energies the temperature dependence of the strange baryons will have similar trends to the 
ones obtained here. The relative increase of $\mu_S(T)$ in our HRG model is comparable to the one deduced from lattice QCD, but 
the absolute values do not match the dependence from lattice QCD~\cite{Borsanyi:2013hza}, which is about 15\% higher than in a standard HRG model. We note that the splitting between the net-pion and net-kaon results is, likewise, a consequence of the 
different strangeness content and the imposed strangeness neutrality.

Preliminary studies, using all measured resonance states listed in PDG-2014~\cite{Agashe:2014kda} up to a mass of $2.5$ GeV/c$^2$, show that
the inclusion of these known higher mass resonances further improves the agreement
with the lattice QCD $\mu_S/\mu_B$~\cite{Alba:publish}. This might suggest the existence of even higher mass, as of yet
unidentified, strange states, some of which are predicted by quark models~\cite{Bazavov:2014xya, Noronha-Hostler:2014aia}, but any further expansion of the resonance mass spectrum
needs to be rigorously tested against all sensitive susceptibility predictions from
lattice QCD.  We have not included those Hagedorn type states in our calculation. We
have verified, though, that the higher $\mu_S$ caused by the additional states listed in PDG-2014
will further increase the temperature dependence
 of the multi-strange baryon $\chi_2$/$\chi_1$ shown in Fig. \ref{fig3} (by $(10-25)\%$ in the relevant temperature range), whereas the particle yield ratios stay rather unaffected.

\begin{figure}[h!]
\centering
\includegraphics[width=0.45\textwidth]{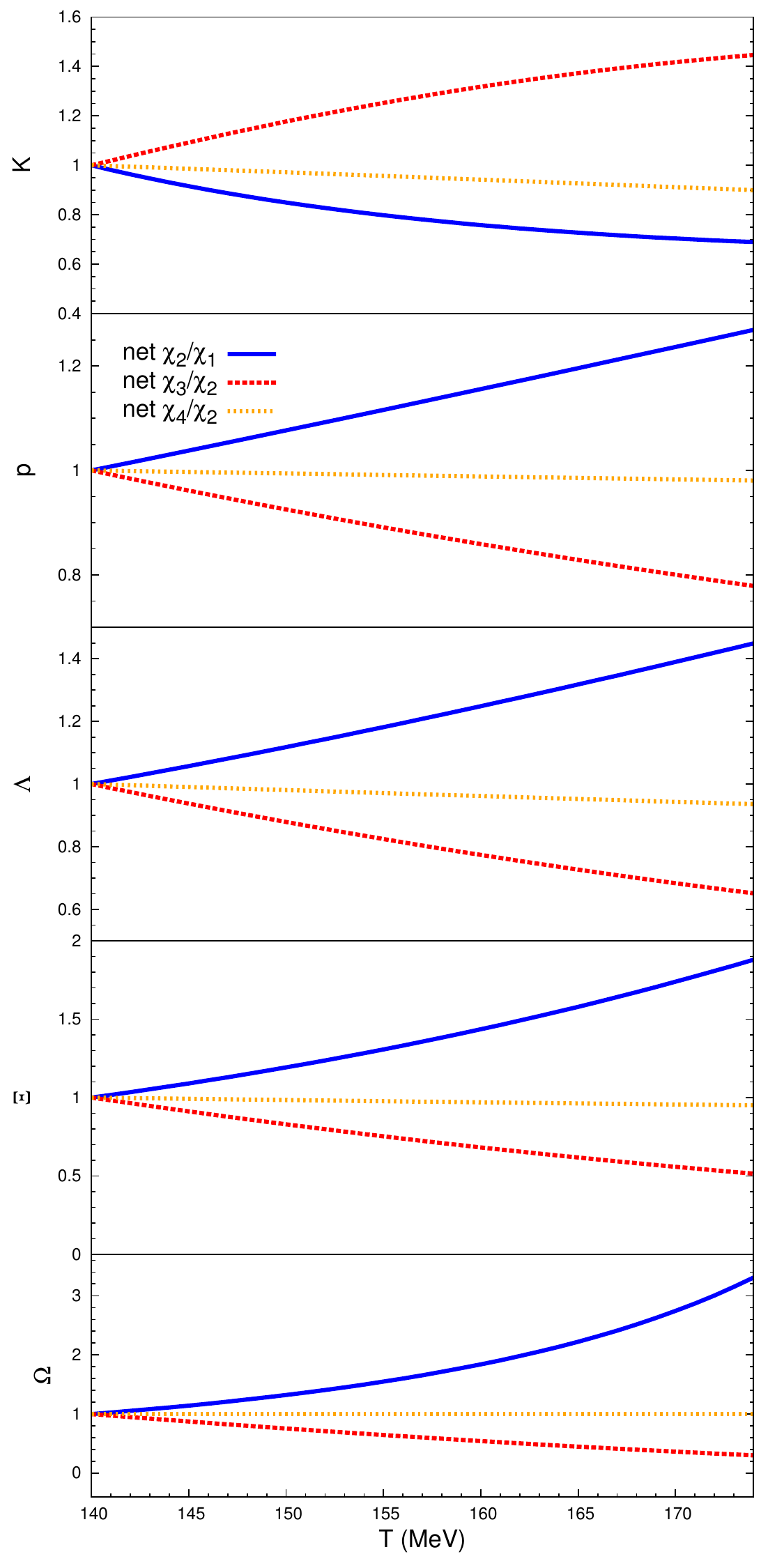}
\caption[]{\label{fig4}(Color online) Comparison of sensitivity of the net-$\chi_2$/$\chi_1$, $\chi_3$/$\chi_2$ and $\chi_4$/$\chi_2$ ratios to the chemical freeze-out temperature. All values are normalized to the proper value at 140 MeV. The baryo-chemical potential has been fixed to 24.3 MeV.}
\end{figure}

Fig. \ref{fig4} shows that the sensitivity of the lower cumulant ratios is also preserved in the $\chi_3$/$\chi_2$ ratio, although 
the sign of the temperature dependence changes. As expected, the kurtosis-based ratio ($\chi_4$/$\chi_2$) shows significantly less sensitivity, 
but its determination is crucial to obtain the strange quark freeze-out conditions, since in lattice QCD, only the even strangeness susceptibilities yield a finite result. 
Therefore any link to first principle calculations needs to be established through $\chi_4$/$\chi_2$, which is the lowest even ratio 
that can be measured. As shown in Fig. \ref{fig4}, each strange particle by itself shows very little temperature dependence, but the more sensitive total net-strangeness $\chi_4$ and $\chi_2$ can be calculated using a combination of all contributing strange particles with pre-factors based on their strangeness content. It should be noted that no PROB corrections were applied to the net-higher moment ratios in Fig.\ref{fig4}.

\begin{figure}[h!]
\centering
\includegraphics[width=0.45\textwidth]{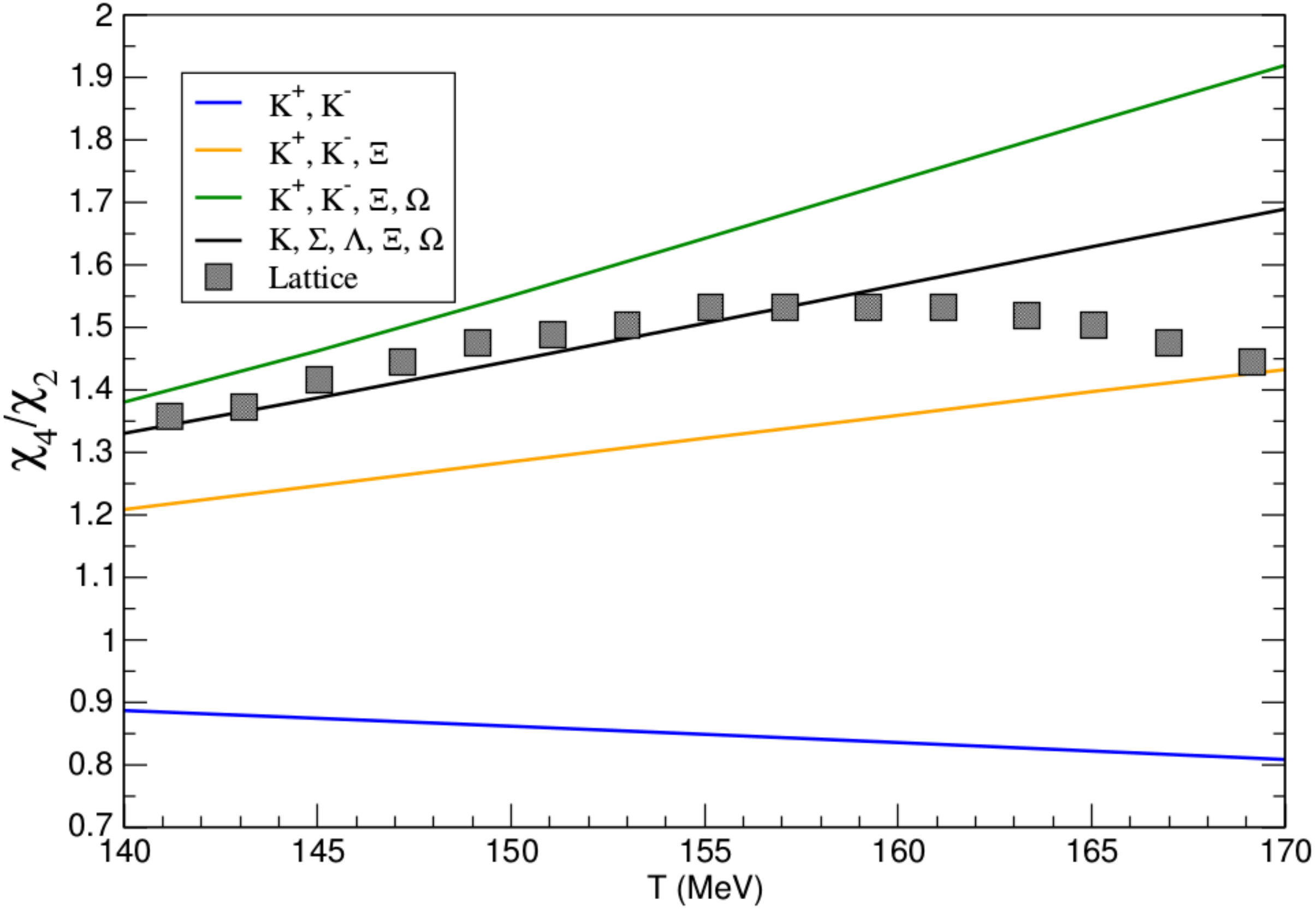}\\
\includegraphics[width=0.45\textwidth]{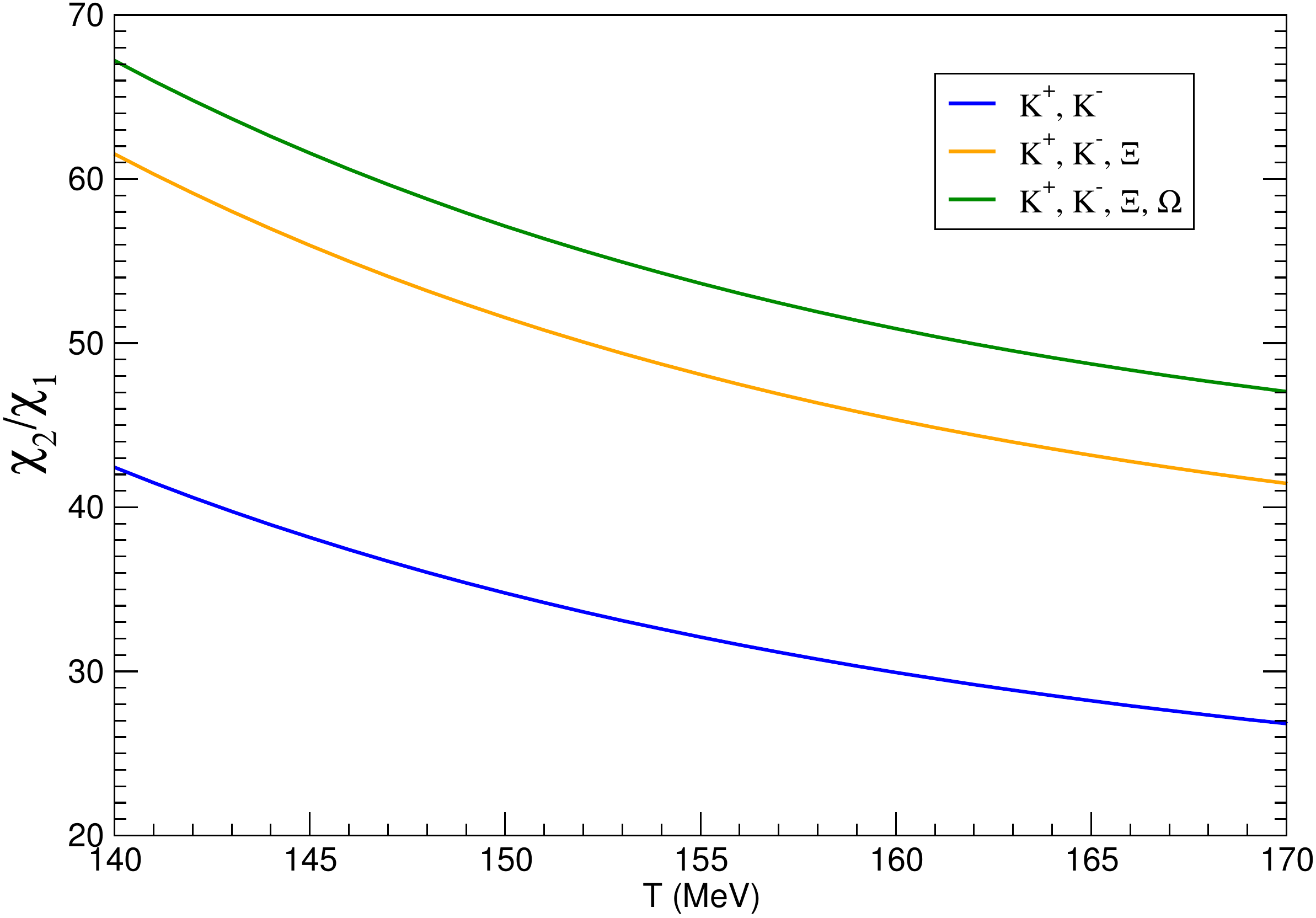}
\caption{\label{fig5}(Color online) Upper panel: comparison of $\chi_4^{net-S}/\chi_2^{net-S}$ obtained from different strange particle combinations. The black 
squares are the lattice QCD results from Ref. \cite{Bellwied:2013cta}. Lower panel: comparison of $\chi_2^{net-S}/\chi_1^{net-S}$ obtained from different strange particle combinations.}
\end{figure}

In Fig.\ref{fig5} we address the question whether measuring a subset of particle species that carry strangeness is sufficient 
to relate the experimental results to calculations of the full strange quark content on the lattice. In the upper panel, we show a comparison
of $\chi_4^{net-S}/\chi_2^{net-S}$ obtained from different strange particle combinations in our HRG model. It is evident that, in order to achieve a 
sensitivity to the temperature, which is comparable to the one on the lattice, multi-strange baryons need to be included in the analysis. Obtaining sufficient statistics in the experiment for higher moments becomes increasingly difficult for such baryons.

A study purely based on a comparison between experimental data and HRG model predictions can be based on $\chi_2$/$\chi_1$ alone, though, since there is no advantage in using the higher cumulants in terms of sensitivity. The lower panel of Fig. \ref{fig5} shows that, contrary to our higher moment study for $\chi_4/\chi_2$, the sensitivity of $\chi_2/\chi_1$ does not benefit from the inclusion of multi-strange baryons: all curves are parallel, thus yielding a comparable sensitivity to $T$. Given the difficulty of obtaining sufficient statistics for multi-strange baryons, it might therefore be even advantageous to look only at lighter strange hadrons as long as one only wants to compare to statistical model predictions. 


\section*{Conclusions}

In this paper we have shown that, for certain final state hadrons, the statistical analysis of the higher moments of the net-particle distributions might add significant sensitivity to the determination of chemical freeze-out parameters compared to a model that uses particle yields and ratios only. In particular, we found that the temperature sensitivity of the net-kaon $\chi_2/\chi_1$ ratio in comparison with the experimentally achievable accuracy is sufficient to reliably determine $T_{ch}$, quite in contrast to the $K/\pi$ ratio. For protons we found a comparable temperature sensitivity to experimental accuracy relation for both, particle and moment ratios.

Our study indicates that it should be possible to obtain a temperature for the chemical freeze-out of strangeness from the efficiency corrected net-kaon $\chi_2/\chi_1$. This might be of particular interest if the strange hadron sector displays a higher $T_{ch}$ than the one extracted  from net-proton and net-charge distributions, which are dominated by light quark particles. For a comprehensive study of the chemical freeze-out properties, we suggest that the experiments provide detailed moment analyses for all identified particle species with sufficient
event-by-event yields. Efficiency corrected experimental results for the lowest four moments are already available for net-protons from the RHIC beam energy scan. Uncorrected data have also been shown for net-kaons, and we are anticipating corrected net-kaon moment distributions from RHIC and LHC in the near future.

\section*{Acknowledgements}

This work is supported by the Italian Ministry of Education, Universities and Research under the Firb Research Grant RBFR0814TT, the US Department of Energy grants DE-FG02-07ER4152, DE-FG02-03ER41260 and DE-FG02-05ER41367 and a fellowship within the Postdoc-Program of the German Academic Exchange Service (DAAD).


\begin{thebibliography}{999}

\bibitem{Cleymans:2005xv}
  J.~Cleymans, H.~Oeschler, K.~Redlich and S.~Wheaton,
  Phys.\ Rev.\ C {\bf 73}, 034905 (2006).
  
\bibitem{Tctrilogy}
  Y.~Aoki {\it et al.},
  Phys.\ Lett.\ B {\bf 643}, 46 (2006);
  Y.~Aoki {\it et al.},
  JHEP {\bf 0906}, 088 (2009);
  S.~Borsanyi {\it et al.}  [Wuppertal-Budapest Coll.],
  JHEP {\bf 1009}, 073 (2010).
  
\bibitem{Aoki:2006we}
  Y.~Aoki, G.~Endrodi, Z.~Fodor, S.~D.~Katz and K.~K.~Szabo,
  Nature {\bf 443}, 675 (2006)  
  
  
  \bibitem{Bazavov:2011nk}
  A.~Bazavov, T.~Bhattacharya, M.~Cheng, C.~DeTar, H.~T.~Ding, S.~Gottlieb, R.~Gupta and P.~Hegde {\it et al.},
  Phys.\ Rev.\ D {\bf 85} (2012) 054503
  [arXiv:1111.1710 [hep-lat]].


  
\bibitem{Sagun:2014sya}
  V.~V.~Sagun, D.~R.~Oliinychenko, K.~A.~Bugaev, J.~Cleymans, A.~I.~Ivanytskyi, I.~N.~Mishustin and E.~G.~Nikonov,
  Ukr.\ J.\ Phys.\  {\bf 59} (2014) 1043
  [arXiv:1403.6311 [hep-ph]].
  

\bibitem{BraunMunzinger:2003zd}
  P.~Braun-Munzinger, K.~Redlich and J.~Stachel,
  In *Hwa, R.C. (ed.) et al.: Quark gluon plasma* 491-599.

\bibitem{Becattini:2003wp}
  F.~Becattini, M.~Gazdzicki, A.~Keranen, J.~Manninen and R.~Stock,
  Phys.\ Rev.\ C {\bf 69}, 024905 (2004)

\bibitem{Becattini:2005xt}
  F.~Becattini, J.~Manninen and M.~Gazdzicki,
  Phys.\ Rev.\ C {\bf 73}, 044905 (2006)

\bibitem{Andronic:2011yq}
  A.~Andronic, P.~Braun-Munzinger, K.~Redlich and J.~Stachel,
  J.\ Phys.\ G {\bf 38}, 124081 (2011).


\bibitem{Karsch:2012wm}
  F.~Karsch,
  Central Eur.\ J.\ Phys.\  {\bf 10}, 1234 (2012).

\bibitem{Bazavov:2012vg}
  A.~Bazavov {\it et al.},
  Phys.\ Rev.\ Lett.\  {\bf 109}, 192302 (2012).

\bibitem{Borsanyi:2013hza}
  S.~Borsanyi, Z.~Fodor, S.~D.~Katz, S.~Krieg, C.~Ratti and K.~K.~Szabo,
  Phys.\ Rev.\ Lett.\  {\bf 111}, 062005 (2013).
  
   \bibitem{Borsanyi:2014ewa}
   S.~Borsanyi, Z.~Fodor, S.~D.~Katz, S.~Krieg, C.~Ratti and K.~K.~Szabo,
   Phys.\ Rev.\ Lett.\  {\bf 113} (2014) 052301
   arXiv:1403.4576 [hep-lat].


\bibitem{Mukherjee:2013lsa}
  S.~Mukherjee and M.~Wagner,
  PoS {\bf CPOD 2013}, 039 (2013).

  \bibitem{Alba:2014eba}
  P.~Alba, W.~Alberico, R.~Bellwied, M.~Bluhm, V.~Mantovani Sarti, M.~Nahrgang and C.~Ratti,
  Phys.\ Lett.\ B {\bf 738} (2014) 305
  [arXiv:1403.4903 [hep-ph]].

\bibitem{Kitazawa:2013bta} 
  M.~Kitazawa, M.~Asakawa and H.~Ono,
  Phys.\ Lett.\ B {\bf 728}, 386 (2014)
  [arXiv:1307.2978].
  
  \bibitem{Sakaida:2014pya} 
  M.~Sakaida, M.~Asakawa and M.~Kitazawa,
  Phys.\ Rev.\ C {\bf 90}, no. 6, 064911 (2014)
  [arXiv:1409.6866 [nucl-th]].

\bibitem{Adamczyk:2013dal}
  L.~Adamczyk {\it et al.}  [STAR Collaboration],
  Phys.\ Rev.\ Lett.\  {\bf 112}, 032302 (2014).

\bibitem{charge}
  L.~Adamczyk {\it et al.}  [STAR Collaboration],
  arXiv:1402.1558 [nucl-ex].
  
  \bibitem{McDonald:2012ts}
  D.~McDonald [STAR Collaboration],
  Nucl.\ Phys.\ A {\bf 904-905}, 907c (2013).
  
\bibitem{Begun:2006jf}
  V.~V.~Begun, M.~I.~Gorenstein, M.~Hauer, V.~P.~Konchakovski and O.~S.~Zozulya,
  Phys.\ Rev.\ C {\bf 74}, 044903 (2006).

\bibitem{Karsch:2010ck}
  F.~Karsch and K.~Redlich,
  Phys.\ Lett.\ B {\bf 695}, 136 (2011).

\bibitem{Fu}
  J.~Fu, Phys.\ Lett.\ B {\bf 722}, 144 (2013).

\bibitem{Garg:2013ata}
  P.~Garg, D.~K.~Mishra, P.~K.~Netrakanti, B.~Mohanty, A.~K.~Mohanty, B.~K.~Singh and N.~Xu,
  Phys.\ Lett.\ B {\bf 726}, 691 (2013).

\bibitem{Nahrgang:2014fza}
  M.~Nahrgang, M.~Bluhm, P.~Alba, R.~Bellwied and C.~Ratti,
  arXiv:1402.1238 [hep-ph].
  
  \bibitem{PDG12} J.~Beringer et al. [Particle Data Group], Phys. Rev. D {\bf 86}, 010001 (2012).
  
\bibitem{Bellwied:2013cta}
  R.~Bellwied, S.~Borsanyi, Z.~Fodor, S.~DKatz and C.~Ratti,
  Phys.\ Rev.\ Lett.\  {\bf 111}, 202302 (2013)
  [arXiv:1305.6297 [hep-lat]].
  
  \bibitem{Alba:2013haa}
  P.~Alba, W.~Alberico, M.~Bluhm, C.~Ratti and R.~Bellwied,
  PoS CPOD {\bf 2013} (2013) 060.
  
  \bibitem{Bluhm:2014lva}
  M.~Bluhm, P.~Alba, W.~M.~Alberico, R.~Bellwied and C.~Ratti,
  J.\ Phys.\ Conf.\ Ser.\  {\bf 509} (2014) 012050.
  
\bibitem{Friman:2011pf}
  B.~Friman, F.~Karsch, K.~Redlich and V.~Skokov,
  Eur.\ Phys.\ J.\ C {\bf 71}, 1694 (2011)
  [arXiv:1103.3511 [hep-ph]].

\bibitem{Andronic:2005yp}
  A.~Andronic, P.~Braun-Munzinger and J.~Stachel,
  Nucl.\ Phys.\ A {\bf 772}, 167 (2006)
  [nucl-th/0511071].

\bibitem{Magestro:2001jz}
  D.~Magestro,
  J.\ Phys.\ G {\bf 28}, 1745 (2002)
  [hep-ph/0112178].
  
\bibitem{Abelev:2008ab}
   B.~I.~Abelev {\it et al.}  [STAR Collaboration],
Au+Au Collisions from STAR,''
   Phys.\ Rev.\ C {\bf 79} (2009) 034909
   [arXiv:0808.2041 [nucl-ex]].
  
  \bibitem{Adams:2006ke}
   J.~Adams {\it et al.}  [STAR Collaboration],
200-GeV,''
   Phys.\ Rev.\ Lett.\  {\bf 98} (2007) 062301
   [nucl-ex/0606014].
 
    \bibitem{McDonald:private}
  D.~McDonald [STAR Collaboration], private communication (2014).
  
\bibitem{Kitazawa:2011wh} 
  M.~Kitazawa and M.~Asakawa,
  Phys.\ Rev.\ C {\bf 85}, 021901 (2012)
  [arXiv:1107.2755 [nucl-th]].
  
  \bibitem{Kitazawa:2012at}
  M.~Kitazawa and M.~Asakawa,
  Phys.\ Rev.\ C {\bf 86} (2012) 024904
   [Erratum-ibid.\ C {\bf 86} (2012) 069902]
  [arXiv:1205.3292 [nucl-th]].  
  
  
  
  
 \bibitem{Agashe:2014kda}
  K.~A.~Olive {\it et al.}  [Particle Data Group Collaboration],
  Chin.\ Phys.\ C {\bf 38} (2014) 090001.
  
    \bibitem{Alba:publish}
  P.~Alba,  R.~Bellwied, V.~Mantovani Sarti and C.~Ratti
  to be published 
  
\bibitem{Bazavov:2014xya}
  A.~Bazavov, H.~-T.~Ding, P.~Hegde, O.~Kaczmarek, F.~Karsch, E.~Laermann, Y.~Maezawa and S.~Mukherjee {\it et al.},
  arXiv:1404.6511 [hep-lat].  
  
    \bibitem{Noronha-Hostler:2014aia}
  J.~Noronha-Hostler and C.~Greiner,
  Nucl.\ Phys.\ A {\bf 931} (2014) 1108
  [arXiv:1408.0761 [nucl-th]].
  
  
  

  
 
  

  

  
  


  

  
  
  




  
  
 


  













  











\end{thebibliography}
\end{document}